\documentclass[]{article}
\usepackage{epsfig}
\catcode`\@=11
\long\def\@makefntext#1{
\protect\noindent \hbox to 3.2pt {\hskip-.9pt  

$^{{\eightrm\@thefnmark}}$\hfil}#1\hfill}		%CAN BE USED 

\def\@makefnmark{\hbox to 0pt{$^{\@thefnmark}$\hss}}	%ORIGINAL 

\def\ps@myheadings{\let\@mkboth\@gobbletwo
\def\@oddhead{\hbox{}
\rightmark\hfil\eightrm\thepage}   

\def\@oddfoot{}\def\@evenhead{\eightrm\thepage\hfil
\leftmark\hbox{}}\def\@evenfoot{}
\def\sectionmark##1{}\def\subsectionmark##1{}}

\newcounter{sectionc}\newcounter{subsectionc}\newcounter{subsubsectionc}
\renewcommand{\section}[1] {\vspace{12pt}\addtocounter{sectionc}{1} 

\setcounter{subsectionc}{0}\setcounter{subsubsectionc}{0}\noindent 

	{\tenbf\thesectionc. #1}\par\vspace{5pt}}
\renewcommand{\subsection}[1] {\vspace{12pt}\addtocounter{subsectionc}{1} 

	\setcounter{subsubsectionc}{0}\noindent 

	{\bf\thesectionc.\thesubsectionc. {\kern1pt \bfit #1}}\par\vspace{5pt}}
\renewcommand{\subsubsection}[1] {\vspace{12pt}\addtocounter{subsubsectionc}{1}
	\noindent{\tenrm\thesectionc.\thesubsectionc.\thesubsubsectionc.
	{\kern1pt \tenit #1}}\par\vspace{5pt}}
\newcommand{\nonumsection}[1] {\vspace{12pt}\noindent{\tenbf #1}
	\par\vspace{5pt}}

\newcounter{appendixc}
\newcounter{subappendixc}[appendixc]
\newcounter{subsubappendixc}[subappendixc]
\renewcommand{\thesubappendixc}{\Alph{appendixc}.\arabic{subappendixc}}
\renewcommand{\thesubsubappendixc}
	{\Alph{appendixc}.\arabic{subappendixc}.\arabic{subsubappendixc}}

\renewcommand{\appendix}[1] {\vspace{12pt}
        \refstepcounter{appendixc}
        \setcounter{figure}{0}
        \setcounter{table}{0}
        \setcounter{lemma}{0}
        \setcounter{theorem}{0}
        \setcounter{corollary}{0}
        \setcounter{definition}{0}
        \setcounter{equation}{0}
        \renewcommand{\thefigure}{\Alph{appendixc}.\arabic{figure}}
        \renewcommand{\thetable}{\Alph{appendixc}.\arabic{table}}
        \renewcommand{\theappendixc}{\Alph{appendixc}}
        \renewcommand{\thelemma}{\Alph{appendixc}.\arabic{lemma}}
        \renewcommand{\thetheorem}{\Alph{appendixc}.\arabic{theorem}}
        \renewcommand{\thedefinition}{\Alph{appendixc}.\arabic{definition}}
        \renewcommand{\thecorollary}{\Alph{appendixc}.\arabic{corollary}}
        \renewcommand{\theequation}{\Alph{appendixc}.\arabic{equation}}
        \noindent{\tenbf Appendix \theappendixc #1}\par\vspace{5pt}}
\newcommand{\subappendix}[1] {\vspace{12pt}
        \refstepcounter{subappendixc}
        \noindent{\bf Appendix \thesubappendixc. {\kern1pt \bfit #1}}
	\par\vspace{5pt}}
\newcommand{\subsubappendix}[1] {\vspace{12pt}
        \refstepcounter{subsubappendixc}
        \noindent{\rm Appendix \thesubsubappendixc. {\kern1pt \tenit #1}}
	\par\vspace{5pt}}

\newcommand{\smalllineskip}{\baselineskip=10pt}

\def\eightcirc{
\begin{picture}(0,0)
\put(4.4,1.8){\circle{6.5}}
\end{picture}}
\def\eightcopyright{\eightcirc\kern2.7pt\hbox{\eightrm c}}

\newcommand{\copyrightheading}[1]
	{\vspace*{-2.5cm}\smalllineskip{\flushleft
	{\footnotesize PAR--LPTHE 96--31}\\
	{\footnotesize hep-th/9609049}\\
	 }}

\def\abstracts#1#2#3{{
	\centering{\begin{minipage}{4.5in}\baselineskip=10pt\footnotesize
	\parindent=0pt #1\par 

	\parindent=15pt #2\par
	\parindent=15pt #3
	\end{minipage}}\par}}

\renewenvironment{thebibliography}[1]
	{\frenchspacing
	 \ninerm\baselineskip=11pt
	 \begin{list}{\arabic{enumi}.}
	{\usecounter{enumi}\setlength{\parsep}{0pt}
	 \setlength{\leftmargin 12.7pt}{\rightmargin 0pt} %FOR 1--9 ITEMS
	 \setlength{\itemsep}{0pt} \settowidth
	{\labelwidth}{#1.}\sloppy}}{\end{list}}

\newcounter{itemlistc}
\newcounter{romanlistc}
\newcounter{alphlistc}
\newcounter{arabiclistc}

\newcommand{\fcaption}[1]{
        \refstepcounter{figure}
        \setbox\@tempboxa = \hbox{\footnotesize Fig.~\thefigure. #1}
        \ifdim \wd\@tempboxa > 5in
           {\begin{center}
        \parbox{5in}{\footnotesize\smalllineskip Fig.~\thefigure. #1}
            \end{center}}
        \else
             {\begin{center}
             {\footnotesize Fig.~\thefigure. #1}
              \end{center}}
        \fi}

\newcommand{\tcaption}[1]{
        \refstepcounter{table}
        \setbox\@tempboxa = \hbox{\footnotesize Table~\thetable. #1}
        \ifdim \wd\@tempboxa > 5in
           {\begin{center}
        \parbox{5in}{\footnotesize\smalllineskip Table~\thetable. #1}
            \end{center}}
        \else
             {\begin{center}
             {\footnotesize Table~\thetable. #1}
              \end{center}}
        \fi}

\def\pmb#1{\setbox0=\hbox{#1}
	\kern-.025em\copy0\kern-\wd0
	\kern.05em\copy0\kern-\wd0
	\kern-.025em\raise.0433em\box0}

\def\fnt#1#2{\footnotetext{\kern-.3em
	{$^{\mbox{\scriptsize #1}}$}{#2}}}

\font\tenrm=cmr10
\font\tenit=cmti10 

\font\tenbf=cmbx10
\font\bfit=cmbxti10 at 10pt
\font\ninerm=cmr9

\font\eightrm=cmr8

\def\qed{\hbox{${\vcenter{\vbox{			%HOLLOW SQUARE
   \hrule height 0.4pt\hbox{\vrule width 0.4pt height 6pt
   \kern5pt\vrule width 0.4pt}\hrule height 0.4pt}}}$}}

\def\jour#1#2#3#4{{#1} {\bf #2}, #3 (#4)}
\def\NPA{{\em Nucl. Phys.} A}
\def\NPB{{\em Nucl. Phys.} B}
\def\PLB{{\em Phys. Lett.}  B}

\def\PRR{{\em Phys. Rev.}}

\def\PRC{{\em Phys. Rev.} C}
\def\PRD{{\em Phys. Rev.} D}

\def\be{\begin{equation}}
\def\ee{\end{equation}}
\def\bea{\begin{eqnarray}}
\def\eea{\end{eqnarray}}
\def\Re{\rm Re}
\begin{document}

\title{{\large \bf On the Physics of a Cool Pion Gas}}
\author{{\small Robert D. Pisarski and 
Michel Tytgat}
\thanks{Talk given by M. Tytgat at the 
``RHIC Summer Studies '96'', Brookhaven 
National Laboratory, New York, USA. mtytgat@wind.phy.bnl.gov}\\
{\small \it Physics Department, Brookhaven National Laboratory,}\\
{\small \it  PO Box 5000, Upton, NY 11973-5000, USA}\\
}
\date{\small September 96}
\maketitle
\abstracts{At finite temperature,  the Nambu-Goldstone bosons of 
a spontaneously
 broken chiral symmetry travel at a velocity $v <
1$. This effect first appears at  order ~$\sim~T^4$ in an 
expansion about low temperature, and can be related to
the appearence of  two  distinct pion decay
constants in a thermal bath. We discuss some consequences on the
thermodynamics of a gas of massless pions.}{}{}

\section{Introduction}
\noindent
In these proceedings, we extend some  previous work of 
ours~\cite{pistyt}. The starting point is very
simple. In the vacuum one invokes Lorentz invariance to define the pion decay
constant, $f_\pi \sim 93 \;{\rm MeV}$, by 
\be
\langle 0 \vert A_\mu^a \vert \pi^b(P) \rangle 
= i f_\pi \delta^{ab} P_\mu \; ,
\label{eq:ea}
\ee
where $A^\mu_a$ is the axial-vector current, and the pion has
euclidean momentum
$P^\mu = (p^0,\vec{p})$. 
At finite temperature, Lorentz invariance is lost and they are
{\em a priori} two
 distinct  pion ``decay constants'': the temporal
component
 has one,
\be
\langle 0 \vert A^{0 a} \vert \pi^b(P) \rangle_T 
= i f^t_\pi \delta^{ab} p^0 \; ,
\label{ft}
\ee
and, assuming O(3) invariance, the spatial part of the current has another,
\be
\langle 0 \vert A^{i a} \vert \pi^b(P) \rangle_T 
= i f^s_\pi \delta^{ab} p^i \; .
\label{fs}
\ee
This is familiar from
nonrelativistic systems, such as discussed by Leutwyler~\cite{leutnon};
in a similar context, this has been recognized by Kirchbach and
Riska~\cite{kirchbach}.  

As in the vacuum, the  pion mass shell is defined using current
conservation, 
\be
\partial^\mu \langle 0\vert A^{\mu a} \vert \pi^b\rangle = 0 \longrightarrow
f_\pi^t \;\omega ^2  = f_\pi^s\; p^2\,.
\label{cons}
\ee
Then, quite trivially,  $f_\pi^t \not= f_\pi^s$
implies  the velocity $v$ \footnote[3]{That $v \leq1$ is required
by causality. It is possible that in some background 
$v>1$, but this is not what our  work is about.}
\be
v^2 = \Re
(f_\pi^s/f_\pi^t) < 1\;.
\label{velo}
\ee
This, again, is 
familiar from other contexts, like  the propagation of light in a
medium -- $v < 1$ corresponds to an index of refraction $n >
1$.  But dealing with Nambu-Goldstone
bosons has some non-trivial
 consequences. For instance, the $f_\pi$'s
develop an imaginary part at finite
temperature. From~(\ref{cons})  one  can  conclude that
the damping rate of massless pions   vanishes at zero
momentum~\cite{pistyt}, an expression of the Goldstone
theorem.\footnote[4]{In a nonlinear $\sigma$ model, $\gamma \sim p
 T^4/f_\pi^4$.}

It is easy to extend these considerations to include explicit
symmetry breaking~\cite{pistyt}. For soft pions, $p \ll f_\pi$, at low
temperature,
$T \ll f_\pi$ -- christened {\em cool pions} in~\cite{pistyt} --  the
dispersion relation is  of the form
\be
\omega^2 = v^2 p^2 + m^2 
\label{disp}
\ee
If $v \not= 1$, there is both a dynamic (position of the pole
in the complex $\omega$ plane, at  $p=0$) and a static mass (position
of the pole in the complex $p$ plane, at $\omega=0$), the two
being related  by 
\be
m_{dyn} = v \cdot m_{stat} \leq m_{stat}
\ee
Incidentally, one can define two Gell-Mann - Oakes - Renner relations:
\be
(\Re f_\pi^{t})^2\; m_{dyn}^2 = 2 m_q\; \langle \bar q q\rangle_T
\ee
or
\be
\Re f_\pi^t\; \Re f_\pi^s\; m_{stat}^2 = 2 m_q\; \langle \bar q q \rangle_T
\ee

A pion  dispersion relation
like~(\ref{disp}) has been particularly advocated  
by Shuryak~\cite{shur} (see also Gale and Kapusta~\cite{gale}), following
 a different line of thought.  Note that $v<1$ implies 
a {\em flattening} of the dispersion
relation at finite temperature. Experimentally such an effect 
might produce  an enhancement  of dileptons from $\pi\pi$ 
annihilations~\cite{gale}. 

\section{Quantitative Results}
\noindent
To  leading order in a low
temperature expansion $\sim T^2/f^2_\pi$, and in the
chiral limit~$m_\pi =0$,
\be 
f_\pi(T) = f_\pi (1 - {T^2/12 f^2_\pi})
\label{leading}
\ee
This result was first obtained, in a different context, by Binetruy
and
 Gaillard~\cite{bine} and subsequently derived by Gasser and
Leutwyler~\cite{gas} using chiral perturbation theory ($\chi$PT). 
Here,~(\ref{leading}) 
implies that  $ f_\pi^t = f_\pi^s$ to leading order. 
This is actually  a consequence of  chiral 
symmetry\footnote[5]{This has been checked recently by 
Bochkarev and
Kapusta~\cite{bochkarev} by comparing  the predictions of 
the $O(N)$ linear and non-linear (using different parametrizations) 
$\sigma$ models.}, as made particularly  
clear by the derivation of Dey, Eletsky and
Ioffe~\cite{dey}.
Using  Current Algebra and PCAC, they showed that 
\be
\langle A_\mu^a A_\nu^b \rangle_T \sim  (1-T^2/6 f_\pi^2)\; 
\langle A_\mu^a A_\nu^b \rangle  +  T^2/6 f_\pi^2 
\;\langle V_\mu^a V_\nu^b \rangle .
\label{theo}
\ee
From pion pole dominance, one
then extracts~(\ref{leading}) from~(\ref{theo}). Besides the mixing with the
vector-vector  correlator, what
is remarkable is that  Lorentz 
invariance is manifest  -- hence that  $v =1$ -- to order $T^2/f_\pi^2$.

Thus, the effect discussed here can only appear at next-to-leading
order~$\sim T^4$. 
In
$\chi$PT this 
implies  computing to two-loop order.  
For the sake of 
the argument,    in~\cite{pistyt}  we instead 
made use of  a  weakly coupled linear 
$\sigma$ model, {\em i.e.} with a
{\em light}  $\sigma$ particle: $m_\sigma^2 = 2 \lambda
\sigma^2$ and $f_\pi \equiv \sigma$, so that $O(T^4/f^2_\pi m_\sigma^2)$
corrections 
dominate over 
the  $O(T^4/f_\pi^4)$ ones.  Expanding in powers of  $T/m_\sigma$, a 
 one-loop calculation 
then suffices to verify~(\ref{velo}).
 The result for the $f_\pi$'s is 
\bea
f_\pi^t &\sim& (1 - t_1 + 3 t_2 + i t_3) f_\pi \\
f_\pi^s &\sim& (1 -t_1 - 5 t_2 - i t_3) f_\pi
\eea
with 
\be
t_1 = T^2/12 f_\pi^2\, ,\;\; t_2 = {\pi^2 T^4\over 45 f_\pi^2
m_\sigma^2}\,,\;\;t_3 = {m_\sigma^4\over 32 \pi f_\pi^2 \omega^2}\;
 \exp(-m_\sigma^2/4 \omega T)
\ee
so that
\be
v^2 \sim 1 - 8 t_2
\ee
which agrees with the direct calculation of the dispersion relation, 
as first carried out  by Itoyama and Mueller~\cite{ito}. 
\vskip 0.2cm

There is no calculation of $f_\pi$ to next-to-leading order in
$\chi$PT, but that $v < 1$ in the chiral limit is implicit
in~\cite{gerber} where Gerber and Leutwyler computed, among other things,
the three-loop corrections to the 
energy
density $u$ of a massless pion gas:
\be
u = {1\over 10} \pi^2 T^4 \left [ 1 + {T^4\over 108 f_\pi^4}\left( 7
\ln{\Lambda_p\over T}-1\right) + {\rm O}(T^6)\right]
\label{u}
\ee 
Apart from the pion decay constant,  the energy
 density depends  on another
scale in the chiral limit, $\Lambda_p~\sim~275$ MeV~\cite{gerber}.
The first term in~(\ref{u}) is the energy density of a
 non-interacting massless
pion gas,
\be
u_0 = 3 \int {d^3 p\over(2 \pi)^3} \; \omega(p) n_B(p) = {1\over 10}
\pi^2 T^4
\label{uo}
\ee
with $\omega(p)^2 = p^2$. Substituting in~(\ref{uo}) a modified 
pion dispersion relation,
\be
\omega^2 = v^2 \; p^2,
\ee
can mimic the effect of the pion interactions. The energy density 
of a gas of  free quasi-pions, 
\be
u = {u_0\over v^{3}},
\ee
reproduces~(\ref{u}),
provided the following estimate of the quasi-pion mean velocity holds:
\be
v \sim 1 - {1\over 3}\;{T^4 \over 108 f_\pi^4}\left ( 7
\ln{\Lambda_p\over T}-1\right)
\ee
In agreement with our previous argument, $v =1$ to O($T^2$). It 
is equivalent\footnote[6]{This is the argument that lead 
Shuryak to~(\ref{disp}).~\cite{shur}} to recognize that there is no  $T^6$ 
correction to the energy
density~(\ref{u}). Also, $v < 1$ for $T < \; \sim 250$ MeV, so that 
 $u > u_0$. 
\vskip 0.2cm

What about massive pions ? The most extensive  work on the
propagation of thermal pions is due  
to Schenk~\cite{schenka,schenkb}. Apparently, he found no evidence  
of~(\ref{disp}). 
However, in  fig.~5 of ref.~\cite{schenka} and 
fig.~7 of ref.~\cite{schenkb},
Schenk plots
$R(p)$, the ratio of the quasiparticle energy, to the pion energy
in free space, as a function of momentum.  
To two loop order, as $p$ increases from zero there is a {\em dip} in $R(p)$:
it first decreases, and then increases,
approaching one from
below.  This is only possible if the quasiparticle energy
$\omega(p)^2 = v^2(p)  p^2 + m_\pi^2(T)$,
with $v(0) < 1$. A.~Schenk, private communication, estimates that 
$v(0) \sim .87$ at $T \sim 150$ MeV.

\section{Outlook}

We conclude with two
remarks. The first  concerns spin waves in
ferromagnets (magnons) and was
brought to our attention by R. Brout.
 The other is on the behavior of $f_\pi$ near the
critical temperature.

In two landmark papers, Dyson~\cite{dyson}, improving on earliers 
ideas of Bloch~\cite{bloch}, developed a formalism to describe the
motion of magnons, and computed 
the low temperature corrections to 
the magnetization. Taking magnon interactions into account, 
he found
\be
M(T)/M(0) \sim   1 - a_0 T^{3/2}  - a_1 T^{5/2}
 - a_2 T^{7/2}  - a_3 T^4
+ {\rm O}(T^{9/2})
\label{dyson}
\ee
Given the magnon dispersion relation
\be
\omega \sim c p^2 + {\rm O}(p^4),
\label{ferro}
\ee
the  $T^{3/2}$  term is the famous prediction of the simple Bloch theory, 
in which  magnons  are treated as
non-interacting bosonic particles. 
The $a_1$ and $a_2$  terms are lattice 
effects (the O($p^4$) terms in~(\ref{ferro})),
and do not concern us. Only the  $a_3$ term
$\sim T^4$ is  due to magnon interactions. 
What is striking is
 that there is no $T^3$ term in~(\ref{dyson}); this 
is reminiscent  of the absence of $T^6$ term in~(\ref{u}). In the
quasi-particle picture, this means that the parameter $c$
in~(\ref{ferro}), is not renormalized to leading order ($T^{3/2}$ in
the present case) but only to next-to-leading order, or $\sim
T^{5/2}$. Also, in both cases -- magnons and
massless pions -- the correction to the ``velocity'' is 
 proportional to the energy density.
 
These are precisely the kind of similarities that an effective
Lagrangian approach,  like $\chi$PT, to the dynamical properties 
of spin waves, both in ferromagnets and antiferromagnets,
 could shed light
on~\cite{leutnon,hof}.
\vskip 0.2cm

Now about $f_\pi$ near $T_c$ ? In a recent paper, Jeon and
Kapusta~\cite{jeon},  computed $f_\pi$ to O($T^2$), both at low $T$ and
near the critical temperature $T_c$, in  an O(N) non-linear 
$\sigma$ model, to
 next-to-leading order in a large N expansion. Here, we give a sketch
of a  
linear $\sigma$ model derivation of their results\footnote[7]{The
Lagrangian can be found  in~\cite{pistyt}.}.

The relevant diagrams
are shown below.
\begin{figure}[hbt]
\centerline{\epsfig{file=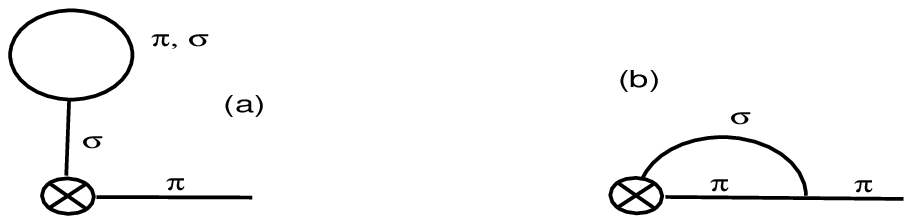}}
\end{figure}
At  low $T \ll m_\sigma$,   $\sigma$ mode propagation 
 is Boltzmann suppressed in diagrams (a) and
(b). The latter shrinks to a tadpole  and  both 
diagrams contribute to order $\sim T^2$. The result, taking into account
wave function renormalization, is
\be
f_\pi(T)= \sigma (1 - T^2/12 \sigma^2),
\ee
 to be compared with the low $T$ dependance of the order
parameter
\be
\sigma(T)= \sigma (1 - T^2/8 \sigma^2)
\ee

Near $T_c$, the $\sigma$ particle is light, $m_\sigma \ll T$,  and,
 in the {\em high}
$T$ expansion, (b) does not contribute nor is there wave
function renormalization to O($T^2$): hence, only the 
tadpole (a) contributes and, 
manifestly,   
\be
f_\pi(T) = \sigma(T) = \sigma ( 1 - T^2/4 \sigma^2)
\label{mf}
\ee
for $T$ near and below $T_c$. 
This agrees with Jeon and Kapusta~\cite{jeon}. That $f_\pi$ and
$\sigma$ vanish at the same temperature is hardly
surprising. An interesting question is whether the critical exponents
of $f_\pi$ and $\sigma$ are  the same also in the critical regime, {\em
i.e.} beyond the mean-field result~(\ref{mf}).

\nonumsection{References}

\end{document}